\documentclass[aps,pra,twocolumn]{revtex4}

\usepackage{rotating}
\usepackage{graphicx}
\usepackage{amssymb}

\begin{document}

\title{X and Y waves in the spatiotemporal Kerr dynamics of a self-guided light beam}
\author{Miguel A. Porras$^1$, Alberto Parola$^2$}
\affiliation{$^1$Departamento de F\'{\i}sica Aplicada, Universidad Polit\'ecnica de Madrid, Rios Rosas 21, ES-28003, Spain\\
             $^2$CNISM and Department of Physics, University of Insubria, Via Valleggio 11, IT-22100
             Como, Italy}
\begin{abstract}
The nonlinear stage of development of the spatiotemporal instability of the monochromatic Townes beam in a medium with
self-focusing nonlinearity and normal dispersion is studied by analytical and numerical means. Small perturbations to the
self-guided light beam are found to grow into two giant, splitting Y pulses featuring shock fronts on opposite sides. Each
shocking pulse amplifies a co-propagating X wave, or dispersion- and diffraction-free linear wave mode of the medium, with
super-broad spectrum.
\end{abstract}

\maketitle

\section{Introduction}

The loud consequences of the instability of the monochromatic Townes beam \cite{CHIAO} in self-focusing Kerr media under
perturbations in space ---catastrophic collapse or diffraction---, as well as the central role played by the Townes profile in
the theory of self-similar collapse \cite{FIBICH,GAETA,SHEN}, have probably overshadowed its relevance in other contexts. The
Townes beam suffers also from an instability under small spatiotemporal perturbations, which was predicted in \cite{ZAKHAROV},
but only studied in detail recently. \cite{PORRAS1} In its initial stages of growth, a linear instability analysis showed that
the spatiotemporal instability is characterized by two weak, Y-shaped unstable modes that can grow exponentially. These weak
Y-shaped modes have been seen to account for the incipient spatiotemporal dynamics observed in a seemingly stationary,
self-guided light beam \cite{PORRAS1} (e.g., a light filament), which includes the onset of spectral broadening and of conical
emission, and the seed for temporal break-up [see for instance Ref. \cite{COUAIRON2} for these phenomena]. In numerical
simulations, weak Y-shaped wave modes have been seen to develop from spatiotemporal perturbations to a Townes beam.
\cite{PORRAS1} In experiments, the single and double Y-shaped structures observed in the spatiotemporal spectra of light
filaments immediately after collapse are compatible with the formation of the unstable modes of the newly formed self-trapped
beam. \cite{PORRAS1}

In this paper we take a step forward by studying the spatiotemporal instability of the Townes beam in its nonlinear regime of
development, where the amplitude of the perturbation becomes comparable to, or higher than the amplitude of the Townes beam, and
the energy lost by the Townes beam is not negligible. Assuming from \cite{PORRAS1} that the Y unstable modes have grown up to a
small amplitude, we show here that they develop into two giant, splitting (subluminal and superluminal) Y pulses that exhibit
broadened spectra and shocks fronts on opposite sides. Moreover, the nonlinear interaction between each Y pulse and the Townes
beam results in the exponential amplification of two X waves with super-broad spectra that co-propagate with the shocking Y
pulses. The Townes beam, the Y pulses and the X waves are seen to coexist during the propagation, interacting strongly in a
highly phase-matched and group-matched configuration. The spatiotemporal dynamics of Y pulses and X waves are found to resemble
very closely the observed dynamics of light filaments, from their formation up to their later stages of proapgation.
\cite{FACCIO,FACCIO2,COUAIRON2,COUAIRON,KOLESIK1,KOLESIK2,BRAGHERI}

Thus, though mathematically complex, the theory on the spatiotemporal instability of a self-guided light beam in \cite{PORRAS1}
(linear regime) and in the present paper (nonlinear regime) can explain from minimum ingredients (diffraction, dispersion and
the Kerr nonlinearity) the origin of the basic phenomena observed in the spatiotemporal dynamics of a self-guided light beam,
though their quantitative description would require, of course, the inclusion of all relevant higher-order effects, as
higher-order dispersion, plasma defocusing, nonlinear absorption, etc.

After recalling briefly the linear instability analysis (Sec. \ref{linear}), we first study by analytical and numerical means
the nonlinear regime of growth of a single Y perturbation at a single frequency to the Townes beam (Sec. \ref{nonlinear}). The
results allows us to understand the most general case of the nonlinear development of the two unstable Y modes when excited at a
continuous of frequencies (Sec. \ref{multiple}).

\section{Linear spatiotemporal instability analysis\label{linear}}

\begin{figure}\begin{center}
\includegraphics[width=3.5cm]{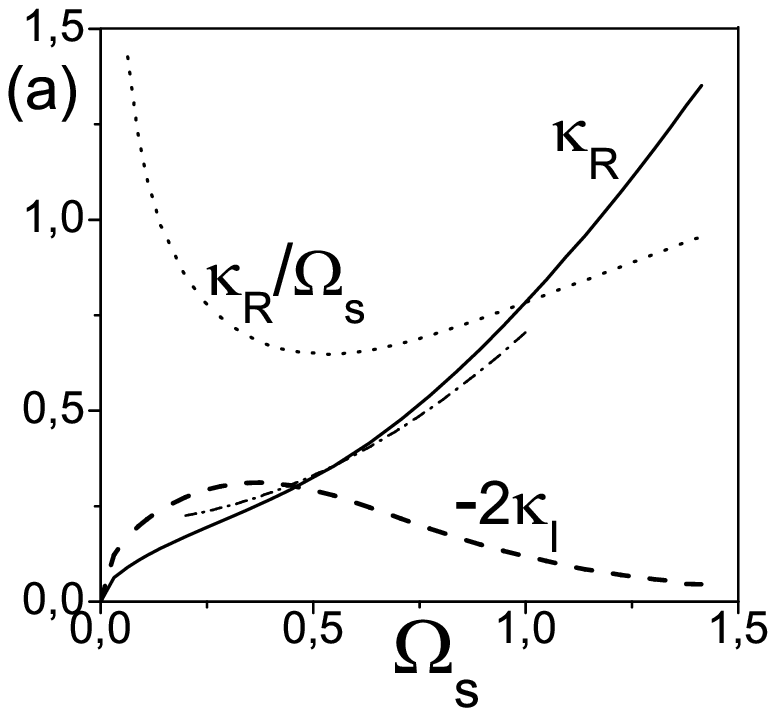}\includegraphics[width=4.2cm]{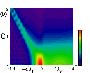}
\caption{\label{fig1} (a) (dashed curve) Gain $-2\kappa_I$, and (solid curve) real part $\kappa_R$ of the eigenvalue  associated
to the unstable modes $p_\kappa$ and $p_{-\kappa^\star}$ at each perturbation frequency $\pm\Omega_s$. (Dotted curve) Quotient
$\kappa_R/\Omega_s$, and (dot-dashed curve) the function $\Omega_s^2/2+\alpha$. (b) Modulus of the spatiotemporal spectrum of
the unstable perturbation $p_\kappa$. Normalization is such that the power $\int_0^\infty|p_\kappa|^2\rho d\rho$ is the same for
all $\Omega_s$.}
\end{center}\end{figure}

We recall that the monochromatic Townes beam $a=a_0(\rho)\exp(i\alpha\xi)$, where $a_0(\rho)$ is the transversal Townes profile,
and $\alpha\simeq 0.2055$ for $a_0(\rho)=1$ \cite{CHIAO}, is a transversally localized, self-guided solution to the cubic
nonlinear Schr\"odinger equation (NLSE), which for a wave $a(\rho,\tau,\xi)$ with revolution symmetry about the propagation
direction $\xi$, reads as
\begin{equation}\label{NLSE}
\partial_\xi a = \frac{i}{2}\frac{\partial_\rho (\rho\partial_\rho a)}{\rho} - \frac{i}{2}\partial^2_\tau a +i |a|^2 a \, .
\end{equation}
The term with the second derivative with respect to the local time $\tau$ accounts for normal group velocity dispersion in the
non-monochromatic case. Apart from the well-known instability of algebraical type under pure spatial perturbations, an
instability under small spatiotemporal perturbations of exponential type has been described recently \cite{PORRAS1}. A couple of
unstable modes of the form
\begin{equation}
p_\kappa=\epsilon[u(\rho)e^{-i\Omega_s\tau+i\kappa\xi}+v^\star(\rho) e^{i\Omega_s\tau-i\kappa^\star\xi}]e^{i\alpha\xi},
\end{equation}
and
\begin{equation}
p_{-\kappa\star}=\epsilon[v^\star(\rho)e^{-i\Omega_s\tau -i\kappa^\star\xi}+u(\rho)e^{i\Omega_s\tau +
i\kappa\xi}]e^{i\alpha\xi},
\end{equation}
with $\kappa_I\equiv \mbox{Im} \kappa<0$, for each temporal frequency shift $\pm\Omega_s$ with respect to the Townes beam
frequency, can grow exponentially with gain $-2\kappa_I$. Figure \ref{fig1}(a) shows the gain $-2\kappa_I$, and
$\kappa_R\equiv\mbox{Re}\kappa>0$, that determines the wave number shift along the $\xi$-direction of the unstable modes with
respect to the wave number of the Townes beam, for each frequency shift $\Omega_s$ of the perturbation. Figure \ref{fig1}(b)
shows the spatiotemporal spectrum of the unstable wave mode $p_\kappa$, obtained by plotting the H\"ankel transforms $u(Q)$ and
$v(Q)$ of $u(\rho)$ and $v(\rho)$ at their respective frequency shifts $+\Omega_s$ and $-\Omega_s$ [similar plot is obtained for
$p_{-\kappa^\star}$ by reflection of Fig. \ref{fig1}(b) about $\Omega_s=0$]. The structure of the spatiotemporal spectrum in the
form of a vaulted Y-letter is due to the definite maxima at radial frequencies
\begin{equation}
Q\simeq \left\{
\begin{array}{ll} 0 & \mbox{for $+\Omega_s$ [i.e., for $u(Q)$], } \\
                                  \sqrt{2}\,|\Omega_s| & \mbox{for $-\Omega_s$ [i.e., for $v(Q)$].} \end{array}\right.
\end{equation}
This structure points out a common origin for the axial and conical radiation of new frequencies observed in self-guided light
beams. Up-shifted new frequencies propagating collinearly with the Townes ($Q\simeq 0$) beam grow jointly with new down-shifted,
non-collinear frequencies ($Q\simeq \sqrt{2}\,|\Omega_s|$) as a consequence of the excitation of the unstable mode $p_\kappa$
(and vice versa for $p_{-\kappa^\star}$). Y-shaped spectra have been shown in \cite{PORRAS1} to develop spontaneously from
perturbations in the Townes beam. In filamentation experiments, single and double Y-shaped spectra have been observed to emerge
soon after collapse. \cite{PORRAS1}

\section{Nonlinear spatiotemporal instability: single Y wave perturbation\label{nonlinear}}

\subsection{Asymptotic analysis}

\begin{figure}\begin{center}
\includegraphics[width=7.5cm]{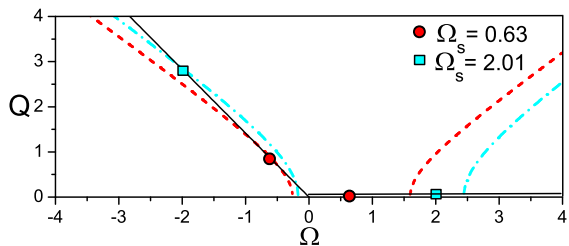}
\caption{\label{fig2} Solid tilted and horizontal lines: Sketch of the Y-shaped $\Omega-Q$ spectrum of $p_\kappa$. Dashed and
dash-dotted curves: X-shaped $\Omega-Q$ spectrum [Eq. (\ref{XSPECTRUM})] for seed frequencies (circles) $\pm\Omega_s=\pm 0.63$
($\kappa_R=0.41$) and for (squares) $\pm\Omega_s=\pm 2.01$ ($\kappa_R=2.23$).}
\end{center}\end{figure}

The validity of the above linear instability analysis is limited to initial stage of growth of the perturbations, where
$|p_\kappa|$, $|p_{-\kappa^\star}|\ll 1$. Some properties of their further evolution can be inferred from the following
analysis. Assume, for simplicity, that only one of the two unstable Y modes, say $p_\kappa$, and at only one couple of
frequencies $\pm\Omega_s$, is growing, as sketched in Fig. \ref{fig2} (circles), starting with small amplitude $\epsilon$. For
small enough $\xi$, the total field $a=\{a_0(\rho)+\epsilon[u(\rho) e^{-i\Omega_s\tau+i\kappa\xi}+v^\star(\rho)
e^{i\Omega_s\tau-i\kappa^\star\xi}]\}e^{i\alpha\xi}$ is seen to satisfy the relation
\begin{equation}\label{MAGIC}
\partial_\xi \bar a= -\kappa_I\,\epsilon\,
\partial_\epsilon\bar a- (\kappa_R/\Omega_s)\,\partial_\tau\bar a  \, ,
\end{equation}
where the reduced field $\bar a= a\exp(-i\alpha\xi)$ is introduced for convenience. Though this relation pertains to the small
perturbation regime, or small enough $\xi$, it is not difficult to demonstrate that it continues to hold along the entire
evolution up to $\xi\rightarrow\infty$ (where the total field continues to depend parametrically on the starting amplitude
$\epsilon$). As said, Eq. (\ref{MAGIC}) is assumed to be satisfied at a certain small value of $\xi$, say $\xi_0$, i.e.,
\begin{equation}\label{MAGIC2}
\partial_\xi \bar a|_{\xi_0}= [-\kappa_I\,\epsilon\,
\partial_\epsilon\bar a- (\kappa_R/\Omega_s)\,\partial_\tau\bar a]|_{\xi_0} \, .
\end{equation}
From the NLSE (\ref{NLSE}), the evolution equation for $\bar a$, valid everywhere, is of the form $\partial_\xi \bar a= {\cal
L}[\bar a] +i |\bar a|^2\bar a$, where ${\cal L}$ is a linear differential operator. The evolution equation for $\bar a$
implies, in particular, that $\partial_\xi(\partial_\xi\bar a)|_{\xi_0}= \partial_\xi[{\cal L}[\bar a]+i|\bar a^2|\bar
a]|_{\xi_0}$. Making use of the linearity of ${\cal L}$ and of relation (\ref{MAGIC2}) at $\xi_0$, we find, after
straightforward algebra,
\begin{equation}\label{MAGIC3}
\partial_\xi(\partial_\xi \bar a)|_{\xi_0}= \partial_\xi[-\kappa_I\,\epsilon\,
\partial_\epsilon\bar a- (\kappa_R/\Omega_s)\,\partial_\tau\bar a]|_{\xi_0} \, .
\end{equation}
The derivatives of the left hand and the right hand sides of Eq. (\ref{MAGIC2}) at $\xi_0$ are also equal. Analogously, starting
from $\partial^n_\xi(\partial_\xi\bar a)|_{\xi_0}=
\partial^n_\xi[{\cal L}[\bar a]+i|\bar a^2|\bar a]|_{\xi_0}$ ($n=2,3\dots$), we also find
\begin{equation}\label{MAGICN}
\partial^n_\xi(\partial_\xi \bar a)|_{\xi_0}= \partial^n_\xi[-\kappa_I\,\epsilon\,
\partial_\epsilon\bar a- (\kappa_R/\Omega_s)\,\partial_\tau\bar a]|_{\xi_0} \, .
\end{equation}
Thus, the full nonlinear evolution governed by the NLSE (\ref{NLSE}) implies that if relation (\ref{MAGIC2}) is valid at
$\xi_0$, it is also valid for all its derivatives at $\xi_0$. This assertion is equivalent to say that relation (\ref{MAGIC2})
is valid not only at $\xi_0$ but also at any propagation distance $\xi$. This last point can be seen in more formal terms by
expressing $\partial_\xi \bar a$ at any distance $\xi$ as the power series
\begin{equation}\label{SERIES}
\partial_\xi \bar a = \sum_{n=0}^{\infty} \frac{1}{n!}\partial^n_\xi(\partial_\xi \bar a)|_{\xi_0} (\xi-\xi_0)^n \, .
\end{equation}
Using Eq. (\ref{MAGICN}) for the successive derivatives at $\xi_0$ in Eq. (\ref{SERIES}), reordering of the different terms, and
adding up, we readily obtain $\partial_\xi \bar a= -\kappa_I\,\epsilon\,
\partial_\epsilon\bar a- (\kappa_R/\Omega_s)\,\partial_\tau\bar a$ at any distance $\xi$, that is, Eq. (\ref{MAGIC}).

From the general solution of (\ref{MAGIC}), the field at any propagation distance is found to be of the form $a=F[\rho;\epsilon
e^{-\kappa_I\xi}; \tau-(\kappa_R/\Omega_s)\xi]\exp(i\alpha\xi)$. In the asymptotic limit $\xi\rightarrow\infty$, $\epsilon
e^{-\kappa_I\xi}\rightarrow\infty$ since $\kappa_I<0$ for an unstable perturbation, and the field must adopt the generic form
\begin{equation}\label{XWAVE}
a=G[\rho;\tau-(\kappa_R/\Omega_s)\xi]\exp(i\alpha\xi)\, .
\end{equation}
If as suggested by the observations, the nonlinear dynamics relaxes into a linear regime at large $\xi$, the field $a$ in
(\ref{XWAVE}) can be identified with an X wave, or a diffraction- and dispersion-free solution to the linear Schr\"odinger
equation in a medium with normal dispersion, \cite{PORRAS2} characterized by a inverse group velocity (or group velocity
mismatch) $\kappa_R/\Omega_s$ with respect to $\tau=0$ (i.e., with respect to a linear plane pulse at the Townes frequency). The
X wave (\ref{XWAVE}) is equivalently described as a superposition of Bessel beams
$J_0(Q\rho)\exp[i(\Omega^2/2-Q^2/2)\xi]\exp(-i\Omega\tau)$ of different frequencies $\Omega$, and transversal wave numbers $Q$
verifying the hyperbolic, or X-shaped, dispersion relation \cite{PORRAS2}
\begin{equation}\label{XSPECTRUM}
Q=\sqrt{2[-\alpha-(\kappa_R/\Omega_s)\Omega + \Omega^2/2]}
\end{equation}
such that the axial wave number $\Omega^2/2-Q^2/2=\alpha+(\kappa_R/\Omega_s)\Omega$ is a linear function of the frequency
$\Omega$.

Figure \ref{fig2} shows two examples (dashed and dot-dashed curves) of the X-shaped $\Omega-Q$ spectra to which perturbed Townes
beams are expected to evolve when the Y perturbations are seeded at two particular couples $\pm \Omega_s$ of frequencies
(circles and squares). The X spectrum presents always a frequency-gap, and does not keep any simple relation of symmetry with
the seeded or Townes frequencies. The left branch of the X spectrum passes always close to the origin and intersects the tilted
Y branch in the vicinity of $-\Omega_s$. The cut-off frequency of the right branch is always higher than $+\Omega_s$,
approaching one to another as $\Omega_s$ increases.

Further algebra shows that the analysis leading to the X spectrum (\ref{XSPECTRUM}) would remain valid for Y perturbations at
multiple couples of frequencies $\pm\Omega_s$ if the quotient $\kappa_R/\Omega_s$ were constant at different perturbation
frequencies $\Omega_s$. Though the value of $\kappa_R/\Omega_s$ obtained from the linear instability analysis is not constant,
it presents a stationary point about the frequencies with maximum gain [see Fig. \ref{fig1}(a), dotted curve], by which it is
reasonable to expect approximate X-like spectra also for multiple Y perturbations, as seen in Section \ref{multiple}.

\subsection{Numerical analysis}

\begin{figure}\begin{center}
\includegraphics[width=4cm]{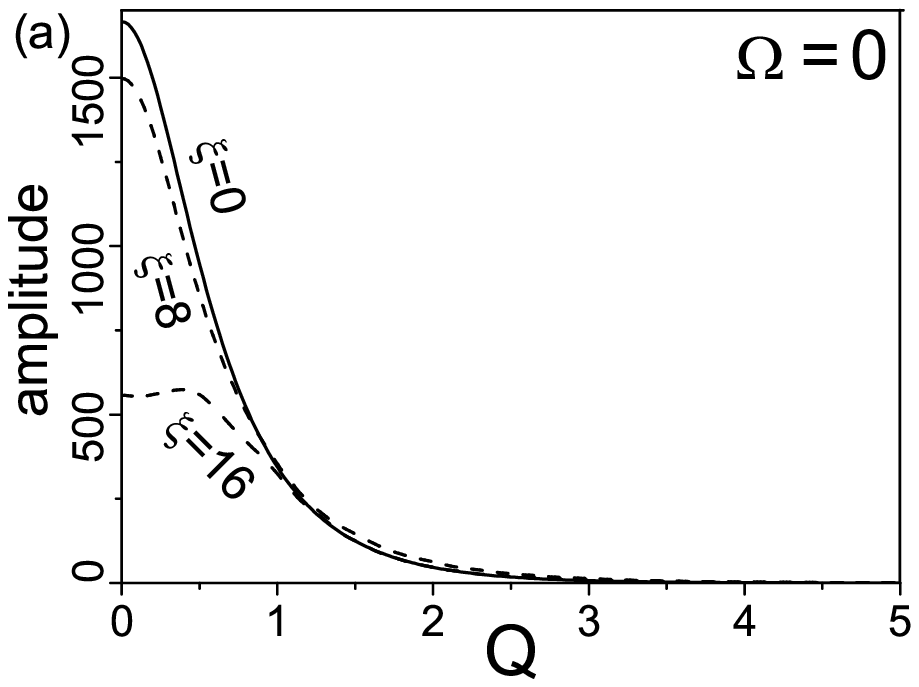}\includegraphics[width=4cm]{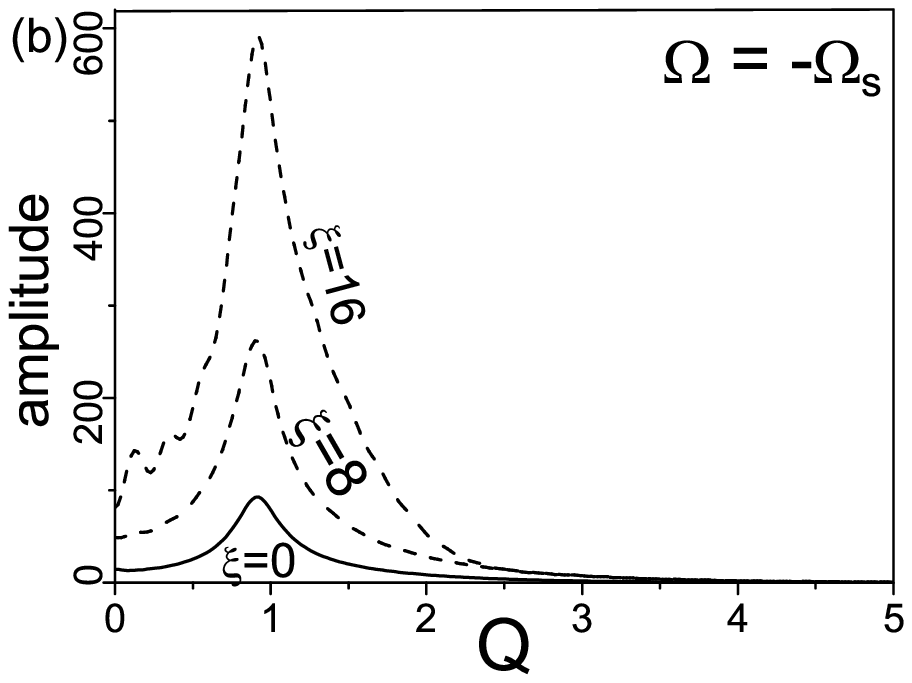}
\includegraphics[width=4cm]{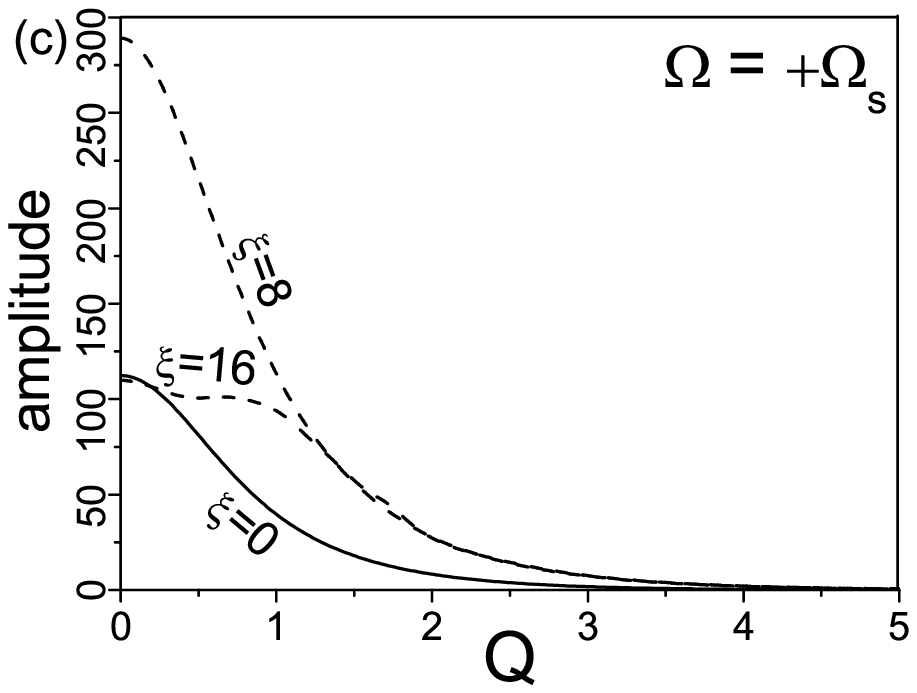}\includegraphics[width=4cm]{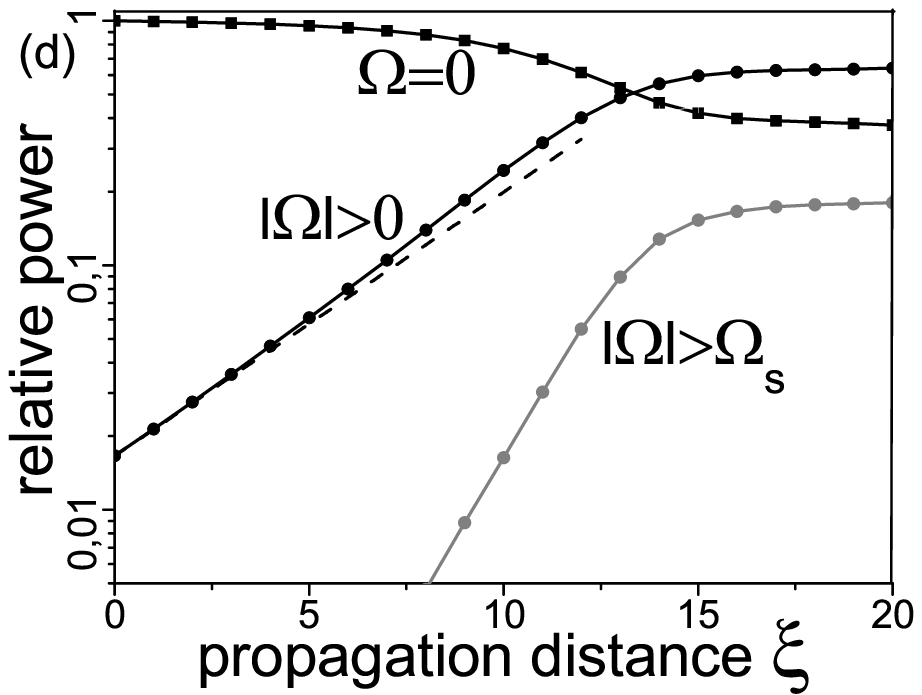}
\caption{\label{fig3} Solid curves: Modulus of the H\"ankel transforms of (a) the Townes profile, and of the Y perturbation (b)
at $-\Omega_s$ and at (c) $+\Omega_s$. The seed $\epsilon$ is such that $|a(\rho,\tau,0)|$ differs from $a_0(\rho)$ by 0.1 as
much, and the perturbation frequency is $\Omega_s=0.63$. Dashed curves: (a) Depletion of the Townes H\"ankel transform, and
growth of the Y perturbation at (b) $-\Omega_s$ and at (c) $+\Omega_s$ upon propagation. (d) In logarithmic scale, total power
in the Townes frequency $\Omega=0$ (squares), in the frequencies $|\Omega|>0$ (circles), and in the newly generated frequencies
$|\Omega|>\Omega_s$ (gray circles) versus propagation distance $\xi$. The dashed straight line represents the exponential gain
$\propto e^{-2\kappa_I\xi}$, with $-2\kappa_I=0.248$ at $\Omega_s=0.63$.}
\end{center}\end{figure}

We have tested the above predictions by numerical integration of the NLSE (\ref{NLSE}) by means of a standard finite-difference
method, taking the perturbed Townes profile
\begin{eqnarray}
a(\rho,\tau,\xi=0)&=&a_0(\rho)+\epsilon[u(\rho) e^{-i\Omega_s\tau}+v^{\star}(\rho) e^{i\Omega_s\tau}]\nonumber\\
                  &=&a_0(\rho)+p_\kappa(\rho,\tau,\xi=0) \label{PERTURBATION}
\end{eqnarray}
as the initial condition. The initial spatiotemporal spectrum $a(Q,\Omega,0)$ of $a(\rho,\tau,0)$ presents then only the three
discrete frequencies $\Omega=0,\pm\Omega_s$ corresponding to the Townes frequency and the Y perturbation $p_\kappa$. Figure
\ref{fig3} shows [(a), solid curve] $a(Q,\Omega\!=\!0,0)\propto a_0(Q)$, or H\"ankel transform of the Townes profile, [(b),
solid curve] $a(Q,\Omega\!=\!-\Omega_s,0)\propto\epsilon v(Q)$, and [(c), solid curve]
$a(Q,\Omega\!=\!+\Omega_s,0)\propto\epsilon u(Q)$, for particular choices of $\Omega_s$ and $\epsilon$ in the perturbation
(\ref{PERTURBATION}) (see caption).

\begin{figure}\begin{center}
\includegraphics[width=3.75cm]{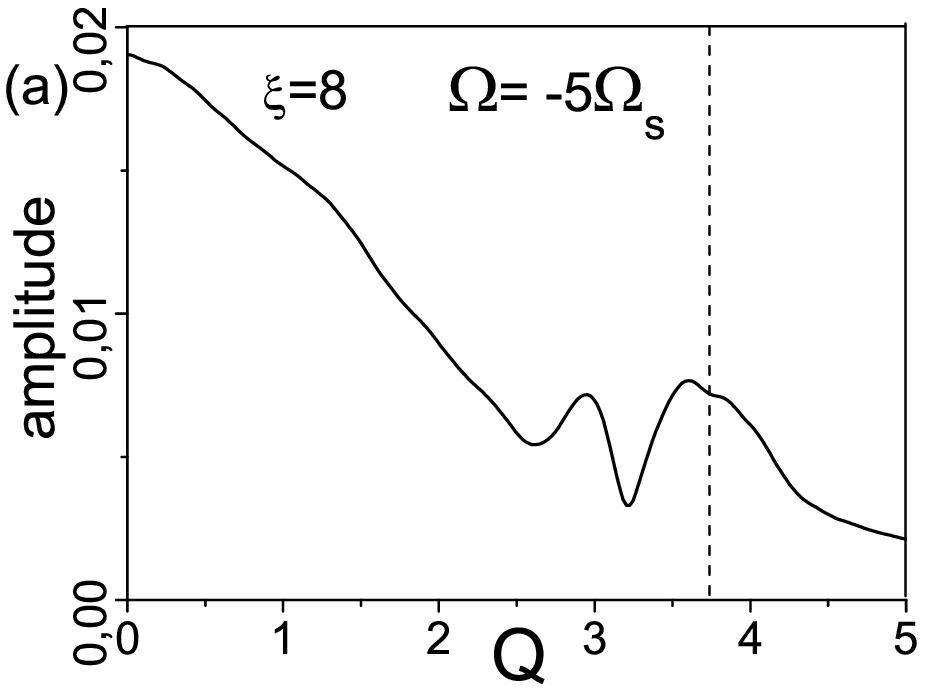}\includegraphics[width=3.75cm]{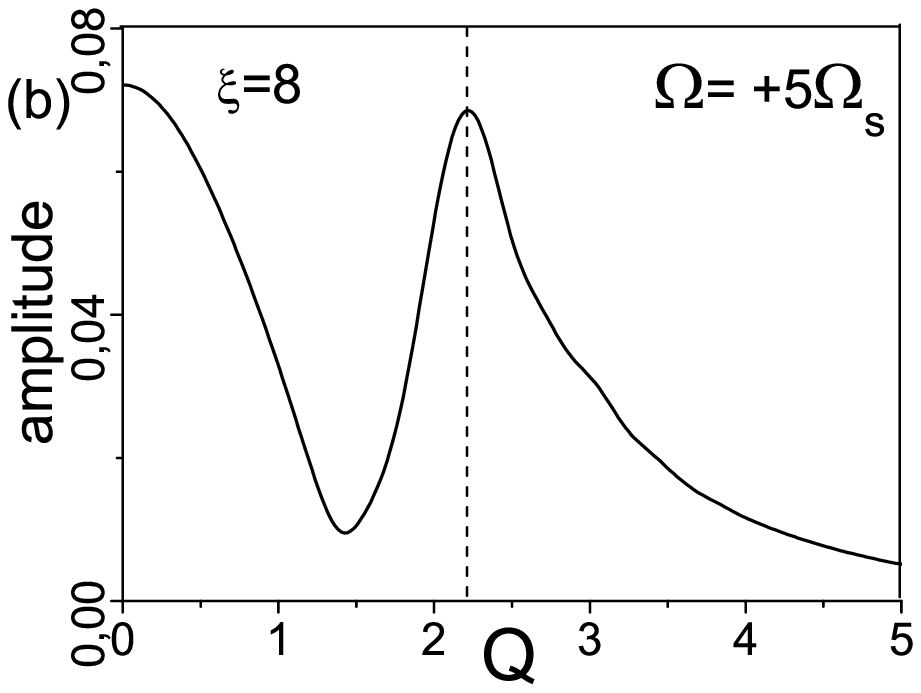}
\includegraphics[width=3.75cm]{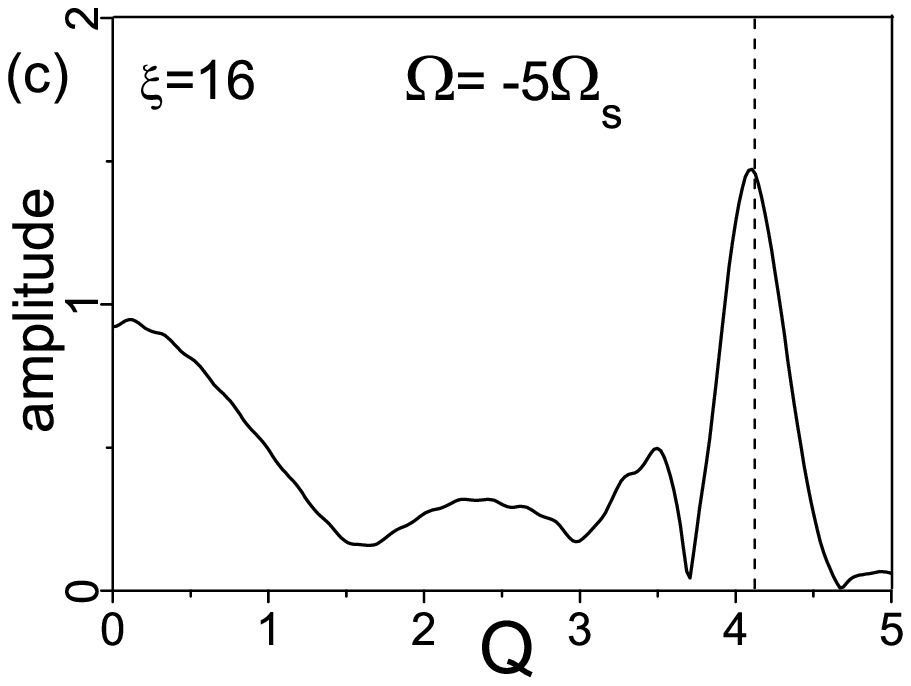}\includegraphics[width=3.75cm]{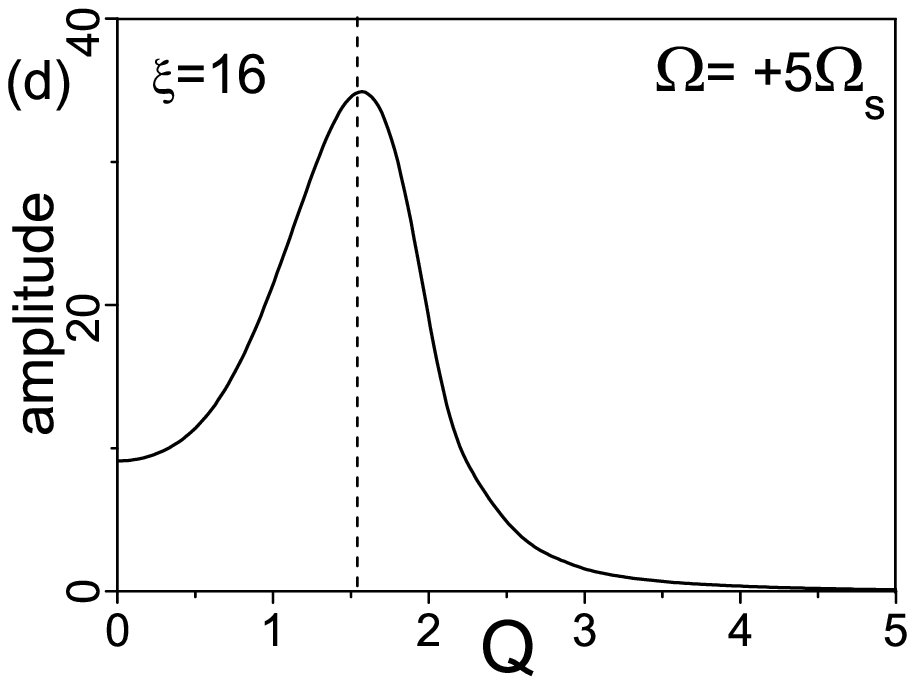}
\includegraphics[width=8cm]{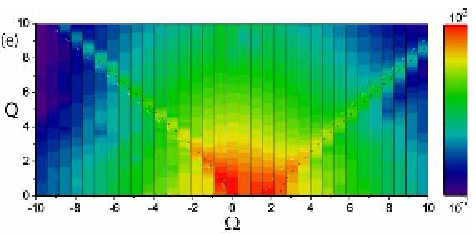}
\caption{\label{fig4} (a) $|a(\Omega=-5\Omega_s,Q,\xi=8)|$ (b) $|a(\Omega=+5\Omega_s,Q,\xi=8)|$, (c)
$|a(\Omega=-5\Omega_s,Q,\xi=16)|$ and (d) $|a(\Omega=+5\Omega_s,Q,\xi=16)|$. The vertical dashed lines are placed at values of
$Q$ given by the X spectrum (\ref{XSPECTRUM}) with perturbation frequency $\Omega_s=0.63$ in (a) and (b), and with perturbation
frequency $\Omega_s=2.01$ in (c) and (d). (e) Contour plot (10 decades in logarithmic scale) of the modulus of the
spatiotemporal spectrum $|a(\Omega,Q,\xi=16)|$. Dashed white curve: X spectrum (\ref{XSPECTRUM}) with $\Omega_s=0.63$. Dashed
black curve: X spectrum with $\Omega_s=2.01$.}
\end{center}\end{figure}

On propagation the power in the Townes frequency $\Omega=0$ decreases [Fig. \ref{fig3}(a), dashed curves] and the perturbation
grows [(b) and (c), dashed curves]. Figure \ref{fig3}(d) displays the power redistribution among the different frequencies
during the propagation. While the power loss in the Townes frequency $\Omega=0$ remains small (closed squares), the power in
frequencies $|\Omega|>0$ (closed circles) grows exponentially with the gain $-2\kappa_I$ (dashed straight line) as expected from
the linear instability analysis. The faster growth of the power in $|\Omega|>0$ observed at longer distances ($\xi\sim 8$ and
beyond) is seen to be due to the generation of additional couples of frequencies $\Omega=\pm N\Omega_s$, $N=2,3,\dots$ with an
exponential gain significantly higher than $-2\kappa_I$ (gray circles). These new frequencies extend over a spectral bandwidth
that exceeds by far the limit expected from the linear instability analysis ($\Omega_s\simeq 1.5$). The spectra of radial
frequencies $Q$ of these new frequencies [see Figs. \ref{fig4}(a) and (b) for the particular choice $\Omega=\pm 5\Omega_s$]
feature a maximum at $Q=0$, which represents a significant enhancement of the axial emission. More importantly, additional
maxima are observed to emerge at a values of $Q$ that fits well, for each new frequency $\Omega$, to the dispersion relation
(\ref{XSPECTRUM}) of the X wave seeded at $\Omega_s$ [vertical lines in Figs. \ref{fig4}(a) and (b)]. Surprisingly, the X wave
(\ref{XSPECTRUM}) is seen to emerge well-before the relaxation of the dynamics with a strong exponential gain.

Another unexpected feature is that the fast exponential growth is accompanied by a slight drift of the X wave spectrum. The weak
maxima of the H\"ankel transforms in Figs. \ref{fig4}(a) and (b) fitting to (\ref{XSPECTRUM}) with $\Omega_s$ (vertical lines)
become gradually shifted to the sharper and much higher maxima in Figs. \ref{fig4}(c) and (d). The shifted maxima fit also to
the dispersion relation (\ref{XSPECTRUM}) of an X wave (vertical lines) but with a perturbation frequency $\Omega_s$ higher than
that in the initial condition. The origin of the drift can be understood from Fig. \ref{fig2}, where we show X wave spectra
(\ref{XSPECTRUM}) at two perturbation frequencies $\pm\Omega_s$, corresponding to the firstly generated X spectrum (dashed
curve) and the final, shifted X spectrum (dot-dashed curve) in the numerical simulation of Figs. \ref{fig3} and \ref{fig4}. For
the Y perturbation at $\pm\Omega_s$ in the initial condition (circles), the lower newly generated frequencies, say
$\pm2\Omega_s$, are characterized by a maximum at $Q\simeq\sqrt{2}|2\Omega_s|$ (enhanced conical emission) for $-2\Omega_s$,
where the left X-branch and Y-branch are nearly indistinguishable, and by a single maximum at $Q=0$ (enhanced axial emission)
for $+2\Omega_s$. No additional maximum appears at $Q\neq 0$ since the frequency $+2\Omega_s$ is lower than the cut-off
frequency of the right X-branch. The couple of new frequencies $\pm2\Omega_s$ then constitutes a secondary Y-perturbation to the
Townes beam with slightly different quotient $\kappa_R(2\Omega_s)/2\Omega_s$, that generates its own, shifted X wave. This
process may be repeated several times during propagation, giving the impression of a drifting X wave spectrum (\ref{XSPECTRUM})
with some average $\Omega_s$ over all Y perturbations.

Relaxation of the nonlinear dynamics occurs when the Townes beam has lost a significant part of its power (more than 50\% in the
example of Figs. \ref{fig3} and \ref{fig4}), and experiences therefore diffraction. As seen in Fig \ref{fig3}(d), interchange of
power among the different frequencies ceases. Figure \ref{fig4}(e) shows the final, asymptotic X-shaped spatiotemporal spectrum
$\Omega$--$Q$, obtained by assembling the H\"ankel transforms at all frequencies $\pm N\Omega_s$ (vertical strips) with
significant power. The most amplified spatiotemporal frequencies lie about either $Q=0$ or the X spectrum (\ref{XSPECTRUM}) with
the shifted $\Omega_s$ (black dashed curve) from the firstly generated X spectrum (\ref{XSPECTRUM}) at the seeded perturbation
frequency $\Omega_s$ (white dashed curve). Extensive numerical simulations shows only quantitative differences in the reported
dynamics for different values of the initial perturbation frequency $\Omega_s$ and its amplitude $\epsilon\ll 1$; no qualitative
changes in the dynamics are observed.

\subsection{Four-wave mixing interpretation}

A simple picture of the above dynamics can be given in terms of the parametric amplification of new spatiotemporal frequencies
in a four-wave mixing interaction driven by the Kerr nonlinearity, as suggested by the exponential growth of the new
frequencies.

The Townes beam is characterized by a frequency $\Omega=0$ and an axial wave number $k_T=\alpha$ due to self-phase modulation
[as in the NLSE (\ref{NLSE}), frequencies, wave numbers and inverse velocities are referred to those of a linear plane pulse at
the Townes frequency]. The Y wave $p_\kappa$ is constituted by the two frequencies $\pm\Omega_s$ with wave numbers
$k_+=\Omega_s^2/2 +2\alpha$ for $+\Omega_s$ (where the first term arises from normal group velocity dispersion and the second
from cross-phase modulation), and $k_-=\Omega_s^2/2-Q^2/2= -\Omega_s^2/2$ for $-\Omega_s$ (where the second term arises from
tilt and the relation $Q=\sqrt{2}|\Omega_s|$ for Y waves has been used). As a wave object, the Y wave itself represents an
infinite train of pulses of carrier frequency $[\Omega_s+(-\Omega_s)]/2=0$, carrier wave number $k_Y=(k_++k_-)/2=\alpha$,
beating frequency $[\Omega_s-(-\Omega_s)]/2=\Omega_s$, and beating wave number $(k_+ - k_-)/2=\Omega_s^2/2 +\alpha$.
Accordingly, the beatings of the Y wave $p_\kappa$ propagate at a inverse group velocity (or group velocity mismatch)
$+\Omega_s/2+\alpha/\Omega_s$.

Once the Y wave has grown up to a significant amplitude, the Townes beam and the Y wave can act as pump waves, and amplify new,
idler and signal waves at frequencies $\pm\Omega$. Note first that a configuration in which the idler and signal frequencies
$\pm\Omega$ form a new Y wave could be efficiently amplified since all Y waves of different frequencies $\pm\Omega$ are
phase-matched with the pump Townes and Y wave. However, the beatings of the new Y wave propagate with different inverse group
velocity $+\Omega/2+\alpha/\Omega$ from those of the pump Y wave at $\pm\Omega_s$.

More efficient amplification is expected for idler and signal frequencies that are not only phase-matched but also group-matched
with the pump waves. For idler wave at $+\Omega$ with $k_1=\Omega^2/2-Q_1^2/2$, signal wave at $-\Omega$ with
$k_2=\Omega^2/2-Q_2^2/2$, producing beatings at inverse group velocity $(-Q_1^2+Q_2^2)/2\Omega$, the condition of phase-matching
($k_T+k_Y=k_1+k_2$) reads as $2\alpha=\Omega^2-(Q_1^2+Q_2^2)/2$, and the condition of group-matching as
$\Omega_s/2+\alpha/\Omega_s=(-Q_1^2+Q_2^2)/2\Omega$. These two conditions lead to
\begin{equation}\label{XSPECTRUM2}
Q_{1,2}= \sqrt{2\left[-\alpha-\left(\frac{\Omega_s^2/2+\alpha}{\Omega_s}\right)\Omega+\frac{\Omega^2}{2}\right]}\, ,
\end{equation}
where $Q_1$ is obtained for positive $\Omega$ and $Q_2$ for negative $\Omega$. Equation (\ref{XSPECTRUM2}) is the same as the
X-shaped spectrum (\ref{XSPECTRUM}), provided that $\kappa_R\simeq \Omega_s^2/2+\alpha$, which in fact has been shown in
\cite{PORRAS1} to approximate $\kappa_R$ where the gain $-2\kappa_I$ is significant [see dot-dashed curve in Fig.
\ref{fig1}(a)].

The X-shaped spectrum is then explained as the result of the preferential amplification of phase- and group-matched
spatiotemporal frequencies by the Y-perturbed Townes beam. In particular, the group velocity of the X wave is the same as the
group velocity of the beatings of the Y wave. Due to the nonsymmetric frequency-gap in the X spectrum, however, the optimum
conditions of phase- and group-matching cannot be attained for $\pm\Omega$ close enough to $\Omega=0$, in which case a new
phase-matched Y wave is amplified.

\section{Nonlinear spatiotemporal instability: double Y perturbations at many frequencies\label{multiple}}

\begin{center}
\begin{figure}
\includegraphics[width=5.5cm]{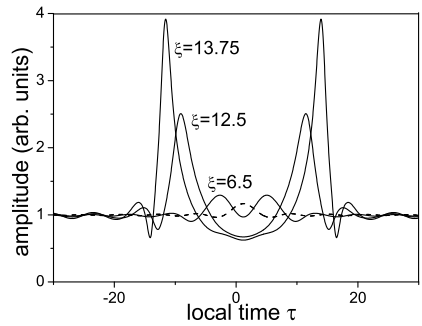}
\caption{\label{fig5} Dashed curve: weakly perturbed Townes beam $|a(\rho=0,\tau,\xi=0)|$ as given by Eq. (\ref{TP2}). The
amplitudes of the two Y pulses are chosen such that $\epsilon_{\kappa,s}=\epsilon_{-\kappa^\star,s}$ and to yield
$|p_{\kappa,s}(0,0,0)|=|p_{-\kappa^\star,s}(0,0,0)|=0.0025$ at all frequencies $\Omega_s$ in the interval $[0,1]$. For this
choice, the two weak Y pulses overlap initially. Solid curves: amplitude $|a(\rho=0,\tau,\xi)|$ at increasing propagation
distances $\xi$.}
\end{figure}
\end{center}

\begin{center}
\begin{figure}[t]
\includegraphics[width=7.5cm]{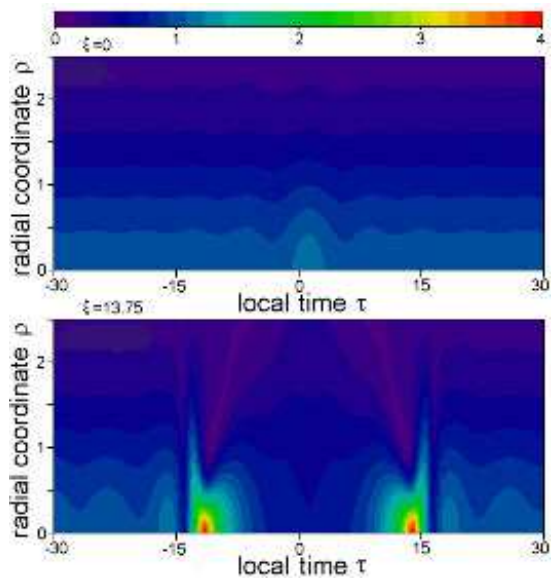}
\caption{\label{fig6} Top: weakly perturbed Townes beam $|a(\rho,\tau,\xi=0)|$ as given by Eq. (\ref{TP2}). Numerical values are
the same as in Fig. \ref{fig5}. Bottom: modulus of the amplitude $|a(\rho,\tau,\xi=13.75)|$ of the propagated field for the
initial condition on the top of the figure.}
\end{figure}
\end{center}

The preceding analysis allows us to understand the more realistic situation in which the self-guided beam experiences
perturbations at a continuous of frequencies $\pm\Omega_s$ about $\Omega=0$. To this purpose we have simulated numerically the
evolution under the NLSE (\ref{NLSE}) of an initial condition of the type
\begin{eqnarray}\label{TP1}
a(\rho,\tau,\xi\!=\!0)&=& a_0(\rho) +\\
              &+& \int_0^\infty\!\! d\Omega_s
\epsilon_{\kappa,s}[u_s(\rho)e^{-i\Omega_s\tau}+v_s^{\star}(\rho)e^{i\Omega_s\tau}] \nonumber \\
                  &+&\!\int_0^\infty\!\! d\Omega_s
\epsilon_{-\kappa^\star,s}[v_s^\star(\rho)e^{-i\Omega_s\tau}\!+ u_s(\rho)e^{i\Omega_s\tau}]\, , \nonumber
\end{eqnarray}
or, equivalently,
\begin{eqnarray}\label{TP2}
a(\rho,\tau,\xi\!=\!0)= a_0(\rho) &+& \int_0^\infty\!\! d\Omega_s p_{\kappa,s}(\rho,\tau,\xi\!=\!0) \\
                          &+& \int_0^\infty\!\! d\Omega_s p_{-\kappa^\star,s}(\rho,\tau,\xi\!=\!0)\, ,\nonumber
\end{eqnarray}
that describes the Townes beam perturbed by the two unstable Y waves excited at a many frequencies $\pm\Omega_s$ with small
amplitudes $\epsilon_{\kappa,s}$ and $\epsilon_{-\kappa^\star,s}$. Each integral in (\ref{TP2}) will be referred to as a weak Y
pulse, since it represents a perturbation of finite duration with the Y-shaped spatiotemporal spectrum of Fig. \ref{fig1}(b)
(and its reflection about $\Omega_s=0$) and small amplitudes determined by $\epsilon_{\kappa,s}$ (and
$\epsilon_{-\kappa^\star,s}$) at each frequency $\pm\Omega_s$. An example of this type of perturbation, with frequencies
$\Omega_s$ filling the interval $[0,1]$ is shown in space-time domain in Figs. \ref{fig5} (dashed curve) at the Townes beam
center $\rho=0$, and in Fig. \ref{fig6} (top) for $\rho\ge 0$ (see captions for numerical values).

Numerical integration of the NLSE (\ref{NLSE}) with a only one weak Y pulse perturbation ($\epsilon_{-\kappa^\star,s}=0$ or
$\epsilon_{\kappa,s}=0$) shows that it develops in the nonlinear regime of instability into a single, steepening, subluminal
pulse (when $\epsilon_{-\kappa^\star,s}=0$) or superluminal pulse (when $\epsilon_{\kappa,s}=0$) of amplitude exceeding the
Townes amplitude. If instead the two weak Y pulses are excited ($\epsilon_{-\kappa^\star,s}\neq 0$ and $\epsilon_{\kappa,s}\neq
0$), two steepening, subluminal and superluminal pulses grow, as seen in Figs. \ref{fig5} (solid curves) and \ref{fig6}
(bottom). These two giant Y pulses can then grow independently and result from the exponential growth of the two weak Y
perturbations in Eqs. (\ref{TP1}) or (\ref{TP2}) beyond the linear regime of instability. Thus, with little loss of generality,
we have limited the numerical simulations to the case $\epsilon_{\kappa,s}=\epsilon_{-\kappa^\star,s}$ leading to symmetric
spectra about $\Omega=0$ and symmetric pulse shapes about $\tau=0$.

\begin{center}
\begin{figure}
\includegraphics[width=3.9cm]{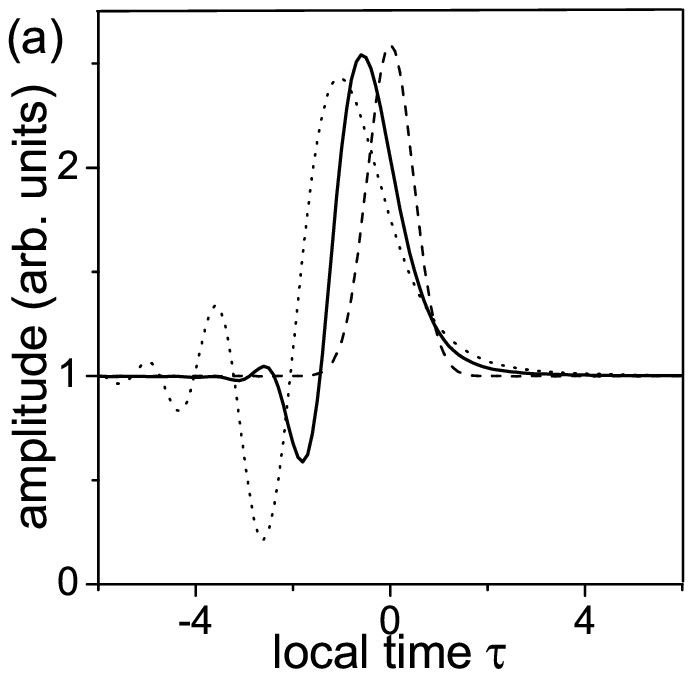}\includegraphics[width=3.9cm]{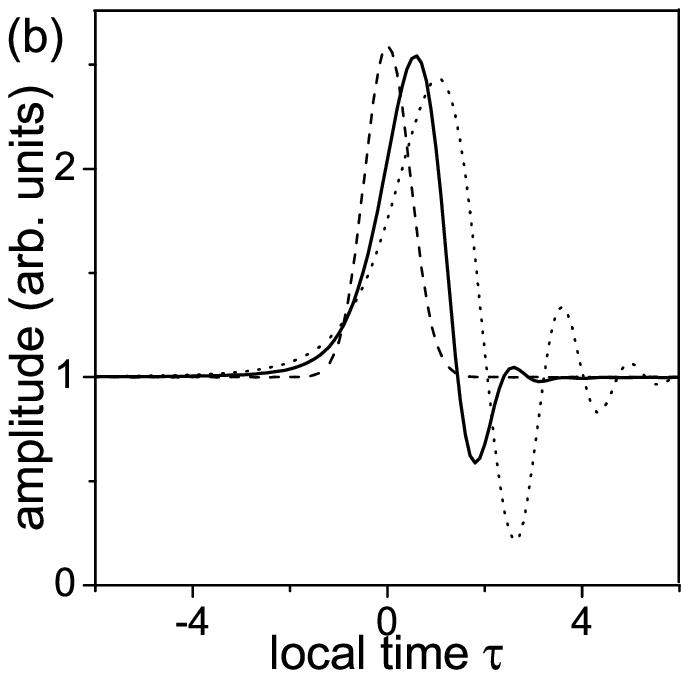}
\caption{\label{fig7} Simplified, one-dimensional model of propagating Y pulses to evidence the opposite group velocities and
steepening in opposite fronts. A pulse at $\xi=0$ (dashed curves) with a Gaussian spectrum [$\exp(-\Omega^2/\Delta\Omega^2)$,
with $\Delta\Omega=3$]  is made to propagate with anomalous (normal) group velocity dispersion for positive (negative)
frequencies in (a), which models $p_{-\kappa^\star,s}$, and vice versa in (b) for $p_{\kappa,s}$. Solid curves: $\xi=0.4$,
Dotted curves: $\xi=0.9$.}
\end{figure}
\end{center}

Subluminality and superluminality, and self-steepening on opposite fronts during propagation, are distinctive features of the Y
pulses, that originate from their reflected Y-shaped spatiotemporal spectra. These features are illustrated in Fig. \ref{fig7}
by means of a simplified model of Y pulses. Contrarily to a standard Gaussian pulse, the positive and negative frequencies in
the spectrum of each Y pulse experience group velocity dispersions of opposite signs. For the Y pulse $p_{\kappa,s}$, e.g., the
frequency $+\Omega_s$ propagates along the $\xi$-direction subject to normal dispersion ($k_+=\Omega_s^2/2+2\alpha$), and the
frequency $-\Omega_s$ does subject to anomalous dispersion ($k_-=-\Omega_s^2/2$), this being a consequence of its tilted
propagation direction. Thus, differently from a Gaussian pulse, opposite portions of the spectrum travel at identical inverse
group velocities along the $\xi$-direction ($dk_+/d\Omega_s|_{\Omega_s}=dk_-/d\Omega_s|_{-\Omega_s}=+\Omega_s$ for
$p_{\kappa,s}$), which prevents from any pulse broadening associated to pulse chirping. In particular, if $\pm\bar\Omega_s$ is a
suitably defined mean frequency of the positive or negative half-spectra of the Y pulse $p_{\kappa,s}$, then its group velocity
as a whole is determined by $+\bar\Omega_s$, i.e., $p_{\kappa,s}$ is subluminal. Similar considerations for the Y pulse
$p_{-\kappa^\star,s}$ lead to superluminality with inverse group velocity $-\bar\Omega_s$. The opposite group velocities of the
two Y pulses can be seen in Fig. \ref{fig7}, where the two halves of a Gaussian-distributed spectrum are forced to propagate
with opposite group velocity dispersions.

Though Y pulses remain unchirped during propagation, they experience broadening and distortion (see Fig. \ref{fig7}) by a
different mechanism from that of standard pulses. Broadening of Y pulses is caused by a chirp in the beatings between each
couple of frequencies $\pm\Omega_s$ due to a dispersion in the beating group velocity, i.e., to a frequency-dependent group
velocity $\Omega_s/2+\alpha/\Omega_s$ for the beatings in the Y pulse $p_{\kappa,s}$, and $-\Omega_s/2-\alpha/\Omega_s$ for
beatings in the Y pulse $p_{-\kappa^\star,s}$. Moreover, assuming a well-behaved, compact spectrum about $\Omega=0$ with mean
positive and negative frequencies $\pm\bar\Omega_s$, the frequencies $\pm 2\bar\Omega_s$ define roughly the extreme frequencies
with significant amplitude in the spectrum, and therefore the fastest beatings in time. They propagate at inverse group
velocities $2\bar\Omega_s/2 + \alpha/2\bar\Omega_s\simeq +\bar\Omega_s$ for $p_{\kappa,s}$, and $-2\bar\Omega_s/2 -
\alpha/2\bar\Omega_s\simeq - \bar\Omega_s$ for $p_{-\kappa^\star,s}$ (where the approximation holds for relatively narrow
spectra, $\bar\Omega_s\gg\sqrt{\alpha/2}=0.32$). In other words, the fastest beatings travel at group velocities approximately
equal to the group velocities of the Y pulses, forming thus the steep fronts about the center of the Y pulses (see Fig.
\ref{fig7}), whose rise times are seen to be directly linked to a frequency $\sim 2\bar\Omega_s$.

\begin{figure}[!t]\begin{center}
\includegraphics[width=6.5cm]{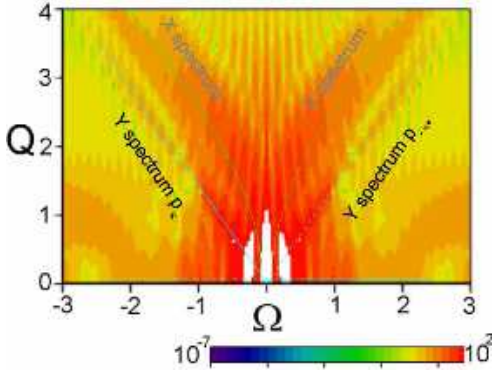}\\
\caption{\label{fig8} In logarithmic scale (9 decades plotted), modulus of the spatiotemporal spectrum $a(Q,\Omega,\xi=16.25)$
of the propagated field $a(\rho,\tau,\xi=16.25)$ for the initial condition in Figs. \ref{fig5} and \ref{fig6}. Solid and dashed
straight lines: double Y spectrum. Solid and dashed curves: double X spectrum.}
\end{center}\end{figure}

The actual shocks in the simulation of Figs. \ref{fig5} and \ref{fig6} show an unbounded increase in the slopes, which can be
explained as the combined effect of Y pulse steepening and of increase of the extreme frequencies, that is, of spectral
broadening in the form of new, phase-matched Y waves at frequencies beyond those in the initial condition, as explained in Sec.
\ref{nonlinear} for single Y wave perturbations. Figure \ref{fig8} displays the hot portion of the spatiotemporal spectrum of
the propagated field for the initial condition in Fig. \ref{fig6} (top), and evidences a significant Y-shaped spectral
broadening. The double Y spectrum, characterized by maxima at the straight lines $Q\simeq\sqrt{2}|\Omega_s|$ and $Q=0$, is
broadened in the propagated field well-beyond the interval $[-1,1]$ in the initial condition. The fine vertical structure
observed in this spectrum originates from interference between the two splitting Y pulses, and the broad maxima above the
Y-shaped spectra are portions of two X-shaped spectra, as described below.

\begin{widetext}

\begin{figure}[!]\begin{center}
\includegraphics[width=17cm]{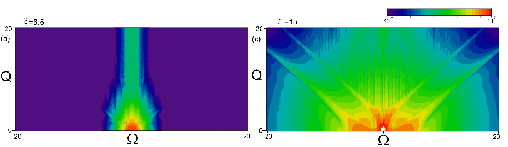}
\hspace*{0.1cm}\includegraphics[width=8.6cm]{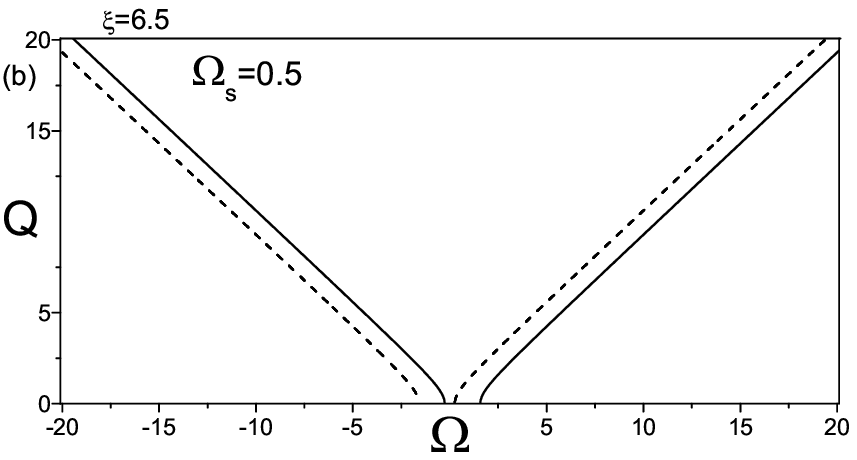}\includegraphics[width=8.6cm]{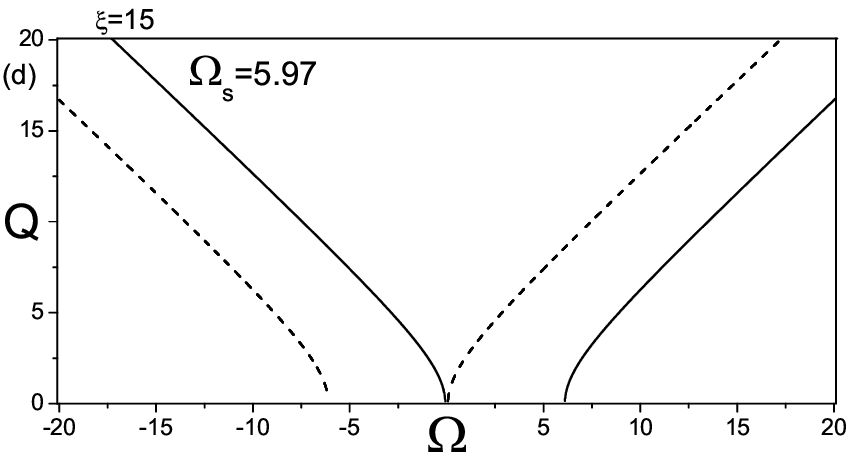}
 \caption{\label{fig9} (a) In logarithmic scale (9 decades plotted), contour plot of modulus of the spatiotemporal spectrum
 $a(Q,\Omega,\xi=6.5)$ for the initial condition in Figs. \ref{fig5} and \ref{fig6},
 where the two Y pulses were seeded at frequencies $\Omega_s $ in $[0,1]$. (b) Fitting double-X spectrum of Eqs. (\ref{DXSPECTRUM1}) and
 (\ref{DXSPECTRUM2}) with
 $\Omega_s=\bar\Omega_s(\xi=6.5)\simeq\Omega_s(\xi=0)= 0.5$ (broadening is very small). (c) In logarithmic scale (9 decades plotted), contour plot of the modulus of the amplitude of the
 spatiotemporal spectrum $a(Q,\Omega,\xi=15)$ for the same initial condition. (d) Fitting double-X spectrum of Eqs. (\ref{DXSPECTRUM1}) and
 (\ref{DXSPECTRUM2}) with
 $\Omega_s=2\bar\Omega_s(\xi\!=\!15)=5.97$.}
\end{center}\end{figure}

\end{widetext}

The nonlinear interaction between each Y pulse and the Townes beam results in the amplification of the two X waves of opposite
group velocity mismatches
\begin{eqnarray}\label{DXSPECTRUM1}
Q_{p_{\kappa,s}}&=&\sqrt{2[-\alpha-(\kappa_R/\Omega_s)\Omega + \Omega^2/2]} \,,\\
Q_{p_{-\kappa^\star,s}}&=&\sqrt{2[-\alpha+(\kappa_R/\Omega_s)\Omega + \Omega^2/2]}\label{DXSPECTRUM2} \,,
\end{eqnarray}
as seen in Fig. \ref{fig8}, or in Fig. \ref{fig9} spanning a much wider spatiotemporal frequency window. The process of
amplification of each X wave from each Y pulse follows the same trend as for single-frequency Y wave perturbations, as expected
from the slow variation of the inverse group velocities, $\kappa_R/\Omega_s\simeq \Omega_s/2+\alpha/\Omega_s$ or
$-\kappa_R/\Omega_s\simeq -\Omega_s/2-\alpha/\Omega_s$, across the spectrum of each Y pulse. Amplification of the two X waves
are independent processes one from another due to the non-overlapping spatiotemporal regions of interaction. In fact, the two
X-shaped spectra in Eqs. (\ref{DXSPECTRUM1}) and (\ref{DXSPECTRUM2}) are seen to develop only when the two Y pulses have
split-off. Figure \ref{fig9}(a) shows an example of the incipient, double-X-shaped spectrum that emerges immediately after Y
pulse splitting (see also Fig. \ref{fig5} at $\xi=6.5$). The location of the maxima of this incipient double-X spectrum is seen
to fit accurately to Eqs. (\ref{DXSPECTRUM1}) and (\ref{DXSPECTRUM2}), as seen in Fig. \ref{fig9}(b), with a perturbation
frequency $\Omega_s$ equal to the mean frequency $\bar\Omega_s(\xi)$ of the half-spectrum, defined by
\begin{equation}
\bar\Omega_s(\xi) \equiv \frac{\int_0^\infty d\Omega \int_0^\infty dQ Q |a(Q,\Omega,\xi)|\Omega}{\int_0^\infty d\Omega
\int_0^\infty dQ Q |a(Q,\Omega,\xi)|}\, .
\end{equation}
With the development of the shock fronts due to the generation of new phase-matched Y waves, the two growing X spectra start to
drift apart, as described for single-frequency Y wave perturbations. Figure \ref{fig9}(c) shows a shifted double-X spectrum at a
distance where the shocks are well-developed, and Fig. \ref{fig9}(d) its fitting to Eqs. (\ref{DXSPECTRUM1}) and
(\ref{DXSPECTRUM2}). Noticeably, the drifting X waves are no longer described by Eqs. (\ref{DXSPECTRUM1}) and
(\ref{DXSPECTRUM2}) with the ``instantaneous" mean frequency $\bar\Omega_s(\xi)$ of the broadening spectrum, as could be
expected, but with $\Omega_s$ close to $2\bar\Omega_s(\xi)$, i.e., twice the instantaneous mean frequency.

This fact can be interpreted to be the result of the preferential amplification of X waves that are highly coupled to the Y
pulses and the self-guided beam. Each X wave is 1) phase-matched with each Y pulse and the Townes beam, as explained in Sec.
\ref{multiple}C, 2) group-matched with the fastest beatings at $2\bar\Omega_s$ in each Y pulse forming its shock, as also
explained in Sec. \ref{multiple}C and in relation to the simple model of Y pulse. In addition, each X wave is 3) group-matched
with each Y pulse, since the group velocity of a Y pulse coincides with the group velocity of its fastest beatings, as seen in
the model of Y pulses.

\section{Conclusions}

Though an accurate description of axial spectral broadening, off-axis spectral broadening or conical emission, pulse splitting,
shock formation, and development double-X-shaped spatiotemporal spectra with frequency-gaps in the propagation of a light
filament, \cite{FACCIO,FACCIO2,COUAIRON2,COUAIRON,KOLESIK1,KOLESIK2,BRAGHERI} requires the inclusion of all relevant, linear and
nonlinear material effects, they have been found to exist with essentially the same characteristics in the spatiotemporal
dynamics of a self-guided light beam in a Kerr medium with normal group velocity dispersion. The key concept unifying all these
phenomena is that of unstable Y-shaped wave modes, that can grow from infinitesimal perturbations to the self-guided beam, and
are seen to develop into two giant, spectrally broadened, splitting and shocking pulses, amplifying two super-broadened, phase-
and group-matched X waves.

In particular, the shock fronts in the trailing edge of the subluminal pulse and in the leading edge of the superluminal pulse,
have been discovered recently in \cite{BRAGHERI}, and have been described as shocking-X waves arising from the subluminal (or
superluminal) propagation and spectral broadening. These shocks appear here as the result of linear Y pulse steepening and
spectral broadening.

Inclusion of higher-order dispersion effects is straightforward and would lead to the so-called ``fish-like" wave modes, or
distorted X waves, in place of the hyperbolic X waves, as observed in experiments and numerical simulations in conditions of
non-negligible high-order dispersion. \cite{KOLESIK1,KOLESIK2,FACCIO3}

In comparison with previous theories of nonlinear X wave (or more generally, fish-like wave) formation, the splitting pulses and
their group velocities are not taken here from numerical simulations of the full Maxwell equations including all material
effects, \cite{KOLESIK1,KOLESIK2} or are not regarded as a source terms with prescribed group velocities in an essentially
linearized Schr\"odinger equation \cite{CONTI}. Splitting with a certain group velocity mismatch appears here as an intrinsic
feature of the unstable Y modes that can grow from (non-splitting) perturbations at noise level. These alternative theories have
been seen to admit also a simple interpretation in terms of a three-wave mixing interaction (rather than four-wave mixing),
\cite{KOLESIK1,KOLESIK2} in which an X wave results from the scattering of an incident wave by a split-off (nonlinear
polarization) pulse, which predicts a linear growth of the X wave, \cite{CONTI} against the exponential and growth obtained from
the present theory. Further experiments could elucidate these points.

M.A. Porras acknowledges Matteo Clerici and Daniele Faccio for helpful discussions, and partial financial support from CNISM
(Consorzio Nazionale Interuniversitario per le Scienze Fisiche della Materia) project for collaborations with researchers from
the European Community.

\end{document}